\documentclass{jfm}
\usepackage{graphicx}
\usepackage{natbib}
\usepackage{amsmath}
\usepackage{amssymb}
\usepackage{units}
\usepackage{upgreek}
\usepackage[normalem]{ulem}

\newlength{\colwidth}
\newcommand{\degrees}{^{\circ}}

\newcommand{\gOne}{\gamma_1}
\newcommand{\gTwo}{\gamma_2}
\newcommand{\gAv}{\overline{\gamma}}				
\newcommand{\etaOne}{\eta_1}
\newcommand{\etaTwo}{\eta_2}
\newcommand{\etaAv}{\overline{\eta}}
\newcommand{\Dg}{\Delta\!\gamma}						
\newcommand{\hN}{h_{\scriptscriptstyle \!N}}						
\newcommand{\hA}{h_{\scriptscriptstyle \!A}}						
\newcommand{\lzero}{l_0}
\newcommand{\ThetaAdjOne}{\Theta_{a,1}}									
\newcommand{\ThetaAdjTwo}{\Theta_{a,2}}									
\newcommand{\ThetaAdjAv}{\overline{\Theta}_a}						
\newcommand{\ThetaAdjAvCrit}{\overline{\Theta}_{a,t}}		
\newcommand{\M}{\widetilde{M}}
\newcommand{\Mcrit}{\M_t}
\newcommand{\Deff}{\widetilde{D}}
\newcommand{\xiA}{\xi_{\scriptscriptstyle\text{Apex}}}
\newcommand{\xiP}{\xi_{\scriptscriptstyle\text{Precursor}}}
\renewcommand{\deg}{^{\circ}}

\newcommand{\abPri}{Pristane}
\newcommand{\abSqA}{Squalane}
\newcommand{\abSqE}{Squalene}

\title[Sharp Transition between Coalescence and Noncoalescence]{Sharp Transition between Coalescence and Noncoalescence of Sessile Drops}
\date{\today}

\author{Stefan Karpitschka and Hans Riegler}
\affiliation{Max Planck Institute of Colloids and Interfaces, 14424 Potsdam, Germany}

\begin{document}
\maketitle

\begin{abstract}

Unexpectedly, under certain conditions, sessile drops from different but completely miscible liquids do \emph{not} always coalesce instantaneously upon contact: the drop bodies remain separated in a temporary state of \emph{noncoalescence}, connected through a thin liquid bridge.
Here we investigate the \emph{transition} between the states of instantaneous coalescence and temporary noncoalescence.
Experiments reveal that it is barely influenced by viscosities and absolute surface tensions.
The main system control parameters for the transition are the arithmetic means of the three-phase angles, $\ThetaAdjAv$ and the surface tension differences $\Dg$ between both liquids. 
These relevant parameters can be combined into a single system parameter, a specific Marangoni number $\M=3\Dg/(2\gAv\ThetaAdjAv^2)$.
This $\M$ universally characterizes the coalescence respectively transition behavior as a function of both, the physicochemical liquid properties and the shape of the liquid body in the contact region.
The transition occurs at a certain threshold value $\Mcrit$ and is sharp within the experimental resolution.
The experimentally observed threshold value of $\Mcrit\approx 2$ agrees quantitatively with values obtained by simulations assuming authentic real space data.
The simulations indicate that the absolute value of $\Mcrit$ very weakly depends on the molecular diffusivity.

\end{abstract}



\section{Introduction}
Sessile drops of identical liquids spreading on a shared substrate will coalesce immediately after coming into contact.
The driving mechanism for drop coalescence is capillarity.
Recently, details of the coalescence behavior of sessile drops have also come into the focus of fundamental research~\citep{Eddi:PhysRevLett111,Castrejon:PhysRevE88,Ristenpart:PhysRevLett97,Hernandez:PhysRevLett109,Borcia:PhysRevE86}.
This is not the least motivated by new applications e.g., ink-jet printing~\citep{Ihnen:ACSApplMatInterf4,Hanyak:JApplPhys109,Stringer:Langmuir26} or microfluidics~\citep{Christopher:LabChip9,Li:LabChip11}.

\begin{figure}
	\centering\includegraphics[width=\colwidth]{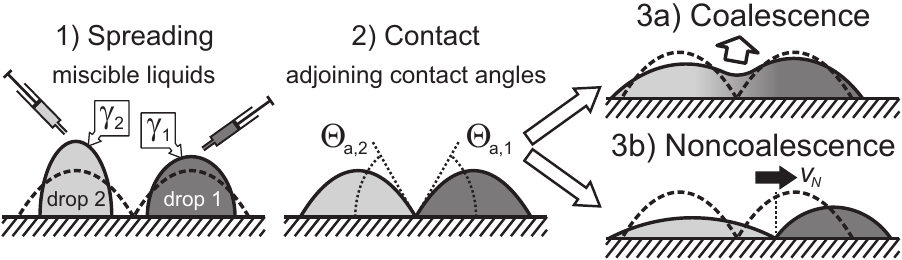}
	\caption{\label{fig:cw:sketch}Sketch of the experimental procedure. Initial volume and distance of the two drops determine the ``adjoining'' three-phase angles, $\ThetaAdjOne$ and $\ThetaAdjTwo$, at which the two drops come into contact. The interplay of contact angles and surface tension contrast leads to coalescence or noncoalescence.}
\end{figure}

In contrast to systems with ``simple'' liquids, much less is known about liquid systems with compositional gradients~\citep{Thiele:PhysRevLett111}.
For example, sessile drops from different but completely miscible liquids with surface tensions $\gOne$ and $\gTwo$ do \emph{not} always coalesce instantaneously upon contact.
After lateral contact, the drop bodies can remain separated in a temporary state of \emph{noncoalescence}, connected through a thin liquid bridge \citep[see Fig.~\ref{fig:cw:sketch} and][]{Riegler:Langmuir24,Karpitschka:Langmuir26,Borcia:Langmuir29}.
Recently this observation, which seemingly contradicts capillarity, has been explained.
The surface tension difference $\Dg = \gamma_1-\gamma_2$ between the drops has been identified as reason for this behavior~\citep{Riegler:Langmuir24,Karpitschka:Langmuir26,Karpitschka:PRL12-2}.
A Marangoni flow induced by $\Dg$ overrides capillarity and temporarily prevents the drop fusion \citep[for an extended qualitative description, see supplemental material of][]{Karpitschka:PRL12-2}. 
Understanding the coalescence behavior of sessile drops from different but completely miscible liquids is important because it represents ubiquitous natural and artificial/technological cases when two bodies of different liquids attached to a solid surface come into contact.
It is for instance relevant to improve a standard semiconductor surface cleaning process\footnote{This research was funded partially by industry (LAM Research AG, Austria)} \citep[``Marangoni drying'', see][]{Leenaars:Langmuir6,Marra:Langmuir7,Matar:PhysFluids13}.
Marangoni flows were also proposed for actuation in microfluidic devices~\citep{Sellier:IntJMultiPhaseFlow37,Sellier:EPJST219}.

The unexpected \emph{state} of temporary noncoalescence has already been analyzed experimentally and theoretically. 
But the \emph{transition} between both modes of coalescence has not yet been investigated up to now.
It is unknown, which parameters control the transition, how sharp it is and which physics cause the transition.
The transition and the related fundamental scientific questions are the subject of this report.


We present the first experimental evidence for a sharp threshold in the coalescence behavior of sessile drops.
As main control parameters we identify the three-phase angles $\Theta$ and the surface tension contrast $\Dg$.
The individual surface tensions ($\gamma_1$, $\gamma_2$) and liquid viscosities ($\eta_1$, $\eta_2$) barely affect the transition. 
The data reveal universal power laws, which couple the topographic ($\Theta$) and the physicochemical properties ($\Dg$) to the coalescence behavior respectively to the transition between both coalescence regimes. 
Within the lubrication approximation~\citep{Oron:RevModPhys69}, we further reduce these power laws to one single, critical Marangoni Number.
Additional numerical simulations reproduce characteristic features of the experimental findings.
They disclose the impact of diffusive mixing on the transition between the two coalescence regimes.

\section{Experimental}%
\begin{figure}
	\centering\includegraphics[width=\colwidth]{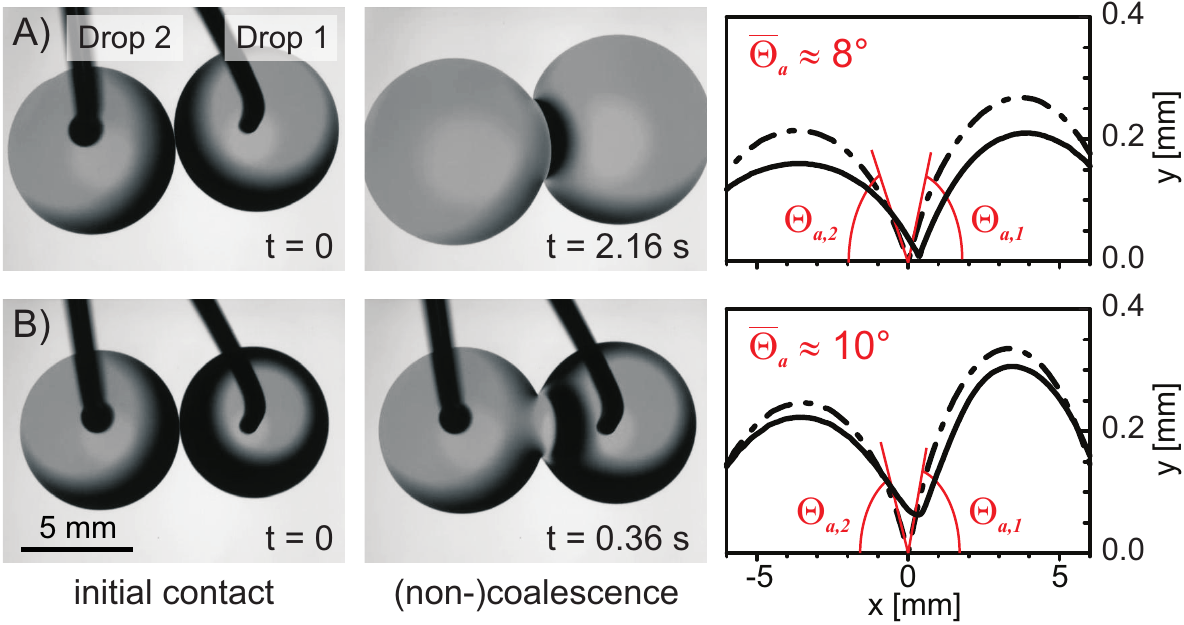}
	\caption{\label{fig:cw:collage}Experiments with tetradecane (drop~2) vs. hexadecane (drop~1). The drops in A) were deposited further apart than in B) leading to smaller adjoining three-phase angles. The result is noncoalescence (A) and immediate coalescence (B). This is revealed by the height profiles through the drop centers (dashed: $t=0$; solid: $t=\unit[2.16]{s}$ and $\unit[0.36]{s}$ for A and B, respectively). For $\ThetaAdjAv\approx\unit[8]{\degrees}$ (A) the connecting neck remains shallow even $\unit[2.16]{s}$ after initial contact, whereas for $\ThetaAdjAv\approx\unit[10]{\degrees}$  (B) the neck height has increased substantially already after $\unit[0.36]{s}$. The syringe tips are not in contact with the drops. See supplement for videos}
\end{figure}

Experiments were performed with completely wetting liquids deposited as separate drops with syringes from the top (Fig.~\ref{fig:cw:sketch}). 
The dynamic three-phase angles $\Theta(t)$ of the continuously spreading drops are used to control the three-phase angles in the moment of mutual drop-drop contact~\citep{Ristenpart:PhysRevLett97,Hernandez:PhysRevLett109}, $\ThetaAdjOne$ and $\ThetaAdjTwo$, the  ``adjoining three-phase angles'' (see Figs.~\ref{fig:cw:sketch},~\ref{fig:cw:collage}).
$\ThetaAdjOne$ and $\ThetaAdjTwo$ depend on the deposition distance/delay, the drop volumes, and the (possibly different) spreading rates.
With this approach identical ($\ThetaAdjOne\approx\ThetaAdjTwo$) as well as different adjoining three-phase angles ($\ThetaAdjOne\neq\ThetaAdjTwo$) can be investigated for identical liquid/liquid- respectively liquid/substrate-combinations (i.e., independently from $\gamma_1$, $\gamma_2$, $\Dg$ and the liquid viscosities $\eta$).

For $mm$ size drops gravitational effects on the initial coalescence behavior are negligible \citep{Ristenpart:PhysRevLett97}
This also holds for variations of the footprint radii/volumes.  

The experiments were performed at $T = \unit[(20.0 \pm 0.5)\degrees]{C}$ in dry nitrogen atmosphere.
The vapor pressures of the liquids are low.
Evaporation effects can be neglected.

As described earlier~\citep{Karpitschka:Langmuir26,Karpitschka:PRL12-2} the coalescence behavior was observed by video imaging from the top and the side. The three-phase angles were measured very accurately by taking both top- and side aspects into account (See supplementary material for details).
The liquids were: a) linear \emph{n}-alkanes ($\text{C}_n\text{H}_{2n+2}$ with $n=13$ to $16$), b) branched alkanes, 2,6,10,14-Tetramethylpentadecane ($\text{C}_{19}\text{H}_{40}$, ``\abPri'') and 2,6,10,15,19,23-Hexamethyltetracosane ($\text{C}_{30}\text{H}_{62}$, ``\abSqA''), and c) a branched alkene, 2,6,10,15,19,23-Hexamethyl-2,6,10,14,18,22-tetracosahexaene ($\text{C}_{30}\text{H}_{50}$, ``\abSqE''). 
The liquids were used as supplied (alkanes: purity $>\unit[99.5]{\%}$, Alfa Aesar; \abPri: purity $\geq\unit[95]{\%}$, Sigma). 
Their surface tensions were measured to check purity.
They agreed with the literature values~\citep{Landolt:IV16}.
The substrates were piranha cleaned silicon wafers (see supplementary material for details). 

The experiments focus on the coalescence behavior immediately after the moment of mutual drop-drop contact. 
The ongoing spreading before and after the contact has no significant impact on the initial coalescence behavior because spreading shortly before contact is typically much slower than the topographical changes right after contact. 


\begin{figure}
	\centering\includegraphics[width=\colwidth]{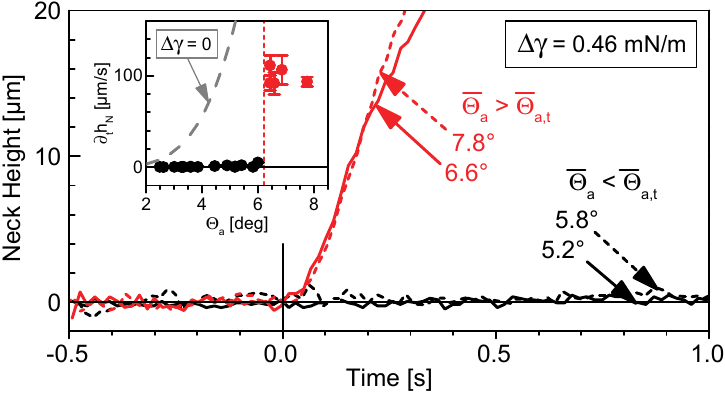}
	\caption{\label{fig:cw:raw}Neck height vs. time (contact at $t=0$) for hexa\-de\-cane vs. pentadecane ($\Dg\approx\unit[0.46]{mN/m}$). The experiments differ only in their $\ThetaAdjAv$ at $t=0$ with a threshold angle $\ThetaAdjAvCrit\approx(\unit[6.2\pm0.4])\degrees$. The noise/height resolution is $\approx\pm\unit[1]{\mu m}$. Inset: Initial neck growth rate vs. $\ThetaAdjAv$ compared to the behavior for identical liquids~\citep{Hernandez:PhysRevLett109} ($\Dg=0$, dashed gray line $\propto \Theta^4$).}
\end{figure}


\section{Experimental results}%
Figures~\ref{fig:cw:collage} A) and B) show the same pair of liquids (tetradecane vs. hexadecane, \hbox{$\Dg\approx\unit[0.87]{mN/m}$}) for different three-phase angles in the moment of contact (``adjoining three-phase angles'' $\Theta_a = \Theta(t\!=\!0)$).
Experiments reveal that the coalescence behavior is insensitive  to the \emph{individual} $\ThetaAdjOne$, $\ThetaAdjTwo$. 
It is only a function of $\ThetaAdjAv=(\ThetaAdjOne+\ThetaAdjTwo)/2$.
The evolution of the drop height profiles for $t>0$ reveals for A) a persistently shallow neck.
For $\ThetaAdjAv\approx\unit[8]{\degrees}$, the coalescence is suppressed (although $\Dg\approx\unit[0.87]{mN/m}$ is remarkably small).
For B) the neck height increases for $t>0$ i.e., for $\ThetaAdjAv\approx\unit[10]{\degrees}$ the coalescence is instantaneous.
This reveals a ``threshold'' angle between temporary noncoalescence and immediate coalescence of $\ThetaAdjAvCrit\approx(\unit[9.0\pm1.0])\degrees$. 

For another case (hexadecane vs. pentadecane, $\Dg\approx\unit[0.46]{mN/m}$) Fig.~\ref{fig:cw:raw} shows the time evolution of the neck heights explicitly. 
The difference between the coalescence behavior is evident with $\ThetaAdjAvCrit\approx(\unit[6.2\pm0.4])\degrees$. 
This smaller $\ThetaAdjAvCrit$ compared to $\ThetaAdjAvCrit$ of the system depicted in Fig.~\ref{fig:cw:collage} correlates with a smaller $\Dg$.
For $\ThetaAdjAv>\ThetaAdjAvCrit$ (immediate coalescence), the Marangoni forces already lead to a decrease in the neck growth rate as compared to identical liquids\citep[see inset and][]{Hernandez:PhysRevLett109}. In contrast to that, for $\ThetaAdjAv<\ThetaAdjAvCrit$, the neck growth is fully suppressed within experimental errors. Therefore we refer to the latter as \emph{temporary noncoalescence}, to avoid confusion with the delaying effect of the Marangoni flow in the regime of immediate coalescence~\citep{Borcia:PhysRevE82}.

\begin{figure}
	\begin{center}%
		\includegraphics[width=\colwidth]{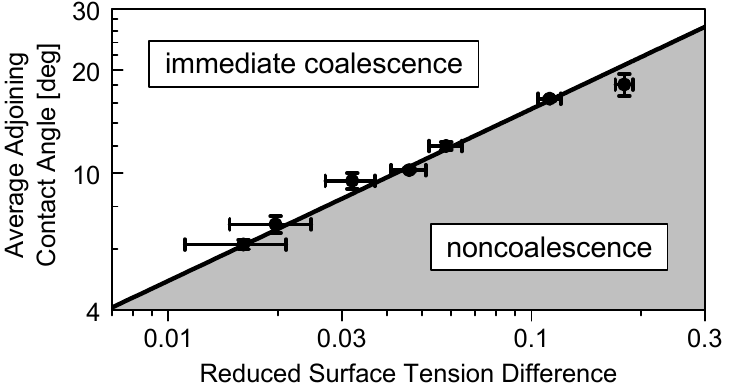}%
	\end{center}
	\caption{
\label{fig:cw:thetacrit}
Threshold behavior of the coalescence of drops from different, miscible liquids: Threshold average adjoining threshold three-phase angle $\ThetaAdjAvCrit$ vs. reduced surface tension difference $\Dg/\gAv$ (double logarithmic plot). Drops with $\ThetaAdjAv>\ThetaAdjAvCrit$ i.e., above the threshold curve, coalesce instantaneously; those with $\ThetaAdjAv>\ThetaAdjAvCrit$ (below the threshold) exhibit noncoalescence. The upper and lower ends of the error bars represent the experiments within each regime that were closest to the border. The line is the result of the scaling (exponent) and the simulations (prefactor), and contains no fit parameter.}
\end{figure}

Fig.~\ref{fig:cw:thetacrit} shows the border between temporary noncoalescence and immediate coalescence as derived from many experiments with various liquid combinations performed as shown in Figures~\ref{fig:cw:sketch},~\ref{fig:cw:collage} and~\ref{fig:cw:raw}. 
The border between both regimes of coalescence is defined by the threshold average adjoining three-phase angle $\ThetaAdjAvCrit$. 
For all investigated combinations of liquids, $\ThetaAdjAvCrit$ (in radians) scales approximately with the square root of $\Dg/\gAv$:
\begin{equation}
	\label{eq:AvthresholdCAexp}
	\ThetaAdjAvCrit\approx \frac{180\deg}{\pi}(0.85\pm 0.05)\,\left(\Dg/\gAv\right)^{0.50\pm0.02}\text{.}
\end{equation}

The liquid viscosities varied by approximately an order of magnitude, from $\unit[2.3]{mPa\,s}$ \citep[Tetradecane,][]{Landolt:IV18} to $\unit[28]{mPa\,s}$ \citep[Squalane,][]{Fermeglia:JChemEngData44}. The surface tensions ranged from $\unit[26.2]{mN/m}$ \citep[Pristane,][]{Korosi:JChemEngData26} to $\unit[31.5]{mN/m}$ \citep[Squalene,][]{Shrestha:Langmuir22}. In agreement with earlier observations on other liquids~\citep{Karpitschka:Langmuir26}, viscosities or absolute surface tensions do not seem to have a significant impact on the threshold behavior. 


\section{Discussion and theoretical analysis}%
The theoretical description shall be reduced to the symmetry ($xz$)-plane of the drops \citep[$z$-direction normal to the substrate, $x$-direction parallel to the substrate through the two apices, compare][]{Ristenpart:PhysRevLett97}.
With drop heights $h<\unit[500]{\upmu m}$, diameters $\unit[2]{mm}<l<\unit[10]{mm}$, and speeds $<{mm/s}$ Reynolds number and aspect ratio are small.
The system can be analyzed in lubrication approximation. 
With a surface tension gradient, the profile evolution equation reads~\citep{Oron:RevModPhys69}:
\begin{equation}
	\label{eq:cw:thinfilm}
	\partial_t h = - \partial_x \left\{\frac{1}{\eta}\left[ \frac{h^3}{3}\partial_x\!\left(\gamma\partial_x^2 h\right) + \frac{h^2}{2} \partial_x\gamma \right]\right\}\text{.}
\end{equation}

The surface tension is described by $\gamma = \gamma_2 + \Dg\,\phi$, with $\gamma_2$ as surface tension of liquid~$2$ and $\phi\in[0,1]$ as local mass fraction of liquid~$1$~\citep{Karpitschka:PRL12-2}. 
$\gamma(x)$ in the capillary term of Eq.~\ref{eq:cw:thinfilm} can be replaced by $\gAv$ because $\Dg\ll\gAv$: Compared to the large, negative capillary pressure in the neck region, the difference in capillary pressure between the two drops' main bodies can be neglected~\citep{Karpitschka:PRL12-2}. The viscosities of the two liquids are similar. In addition, experiments show that the difference of the viscosities does not measurably influence the transition between the coalescence modes~\citep{Karpitschka:Langmuir26}. Thus we replace $\eta(x)$ by $\etaAv=(\etaOne+\etaTwo)/2$.
To account for the low aspect ratio $\epsilon$ of the profiles, we scale $\xi = h/\lzero$ and $\chi = \epsilon x/\lzero$, where $\lzero$ is the (vertical) length scale of the profile. We define $l_0$ via the drops' apex heights in the moment of mutual contact (see below). The scale of the slopes in the profile at mutual drop-drop contact is given by $\ThetaAdjAv$.
Therefore we set the aspect ratio:
\begin{equation}
	\epsilon = \ThetaAdjAv \cdot \pi / 180\deg\text{.}
\end{equation}
Thus, in dimensionless units, the slope $\partial_{\chi}\xi\lessapprox 1$.
The time variable is scaled as $\tau = v_c t/\lzero$, with a capillary velocity $v_c=\gAv\epsilon^4/(3\eta)$~\citep{Aarts:Science7}.
The evolution equation becomes
\begin{equation}
	\label{eq:cw:thinfilmscaled}
	\partial_{\tau} \xi = - \partial_{\chi}\big[ \underbrace{\xi^3\partial_{\chi}^3 \xi}_{=\Phi_C} + \underbrace{\M \xi^2\partial_{\chi} \phi}_{=\Phi_{M}} \big] \text{,}
\end{equation}
with an effective Marangoni number
\begin{equation}
	\label{eq:scaledMarangoninumber}
	\M=3\Dg/(2\gAv\epsilon^2)\text{.}
\end{equation}

$\Phi_C$ in Eq.~\ref{eq:cw:thinfilmscaled} is the flow due to capillarity \emph{from both sides} into the neck region.
It changes sign at the neck \emph{regardless} of the individual curvatures / surface tensions of the two main drop bodies, because the neck region is curved opposite and much stronger as compared to the main drop bodies.
$\Phi_M$ is the Marangoni flow due to the (local) surface tension gradient. $\Phi_M>0$ everywhere, since $\partial_{\chi}\phi>0$ everywhere.
Thus, $\Phi_M$ can compensate for $\Phi_C$ i.e., suppress coalescence, if $\Phi_M$ is strongest / localized to the region where $\Phi_C<0$.
As long as $\Phi_M$ can compensate for $\Phi_C$, coalescence is suppressed and the drops form a traveling wave~\citep{Karpitschka:PRL12-2}.
If $\Phi_M$ cannot compensate $\Phi_C$, immediate coalescence occurs.
$\Phi_M$ scales with $\M \propto \Dg/(\gAv\epsilon^2)$, compared to $\Phi_C$. 

Since the typical neck heights during noncoalescence are in the range of $\unit{\upmu m}$ \citep[for immediate coalescence, they quickly grow to much larger values, compare][]{Karpitschka:PRL12-2,Hernandez:PhysRevLett109}, we do not include a disjoining pressure \citep[which acts for \protect{$h\lessapprox\unit[10]{nm}$},][]{Israelachvili} in our analysis. For the case here (completely wetting liquids), the disjoining pressure, in addition to capillarity, promotes coalescence during the very first moments of drop-drop contact.

Associating the threshold behavior in coalescence with a universal threshold Marangoni number, $\Mcrit$,
Eq.~\ref{eq:scaledMarangoninumber} infers that $\ThetaAdjAvCrit$ scales with: 
\begin{equation}
	\label{eq:AvthresholdCAtheo}
	 \ThetaAdjAvCrit = \frac{180\deg}{\pi}\sqrt{\frac{3\Dg}{2\Mcrit\gAv}}\text{.}
\end{equation}
This agrees with the experimental data (Eq.~\ref{eq:AvthresholdCAexp}) yielding:
\begin{equation}
	\Mcrit \approx 2.0\pm 0.2\text{.}
\end{equation}
$\Mcrit$ marks the shift in superiority between the two lubrication flow contributions that exist if there is a (local) surface tension gradient characterized by $\Dg$ and a non-planar liquid surface characterized by $\ThetaAdjAv$. The result of the competing contributions is either separation of the liquid bodies (Marangoni dominates, $\M>\Mcrit$) or their convergence (capillarity dominates, $\M<\Mcrit$).


\section{Simulations}%
To gain additional insights we also performed numerical simulations.
These focused in particular on $\Mcrit$, its relation to the molecular diffusivity of liquid 2 in liquid 1, $D$, and the category of the transition characterized by $\Mcrit$.
$D$ affects the local surface tension gradient via the local composition evolution~\citep{Karpitschka:PRL12-2}.
Thus it influences the coalescence behavior.
Experimentally $D$ can hardly be varied independently from other system parameters.
Therefore, simulations with varying $D$ are particularly useful.

We analyzed numerically Eq.~\ref{eq:cw:thinfilmscaled} coupled to the evolution equation for $\phi$ (the vertically averaged mass fraction of liquid~1) in lubrication approximation~\citep{Oron:RevModPhys69}:
\begin{equation}
	\label{eq:cw:concevo}
	\partial_{\tau} (\phi\,\xi) = - \partial_{\chi}\big\{\phi\big[ \xi^3\partial_{\chi}^3 \xi + \M \xi^2\partial_{\chi} \phi \big] - \Deff \xi \partial_{\chi}\phi \big\}\text{,}
\end{equation}
where $\Deff=3\eta D/(\lzero\gAv\epsilon^2)$ is the scaled diffusion constant.

Except for $\Deff$, the length scale $\lzero$ does not appear in the equations or the remaining (dimensionless) constants. Thus, the size of the drops in the simulation can be matched to any drop size in the experiments (as long as they are small enough to neglect gravity compared to capillarity), by defining $\lzero$ appropriately. $\Deff$ was chosen adequately to yield realistic values for $D$ after defining $\lzero$. In addition, $\Deff$ was varied by about an order of magnitude around that value.

The simulations started with two parabola as ``drops'' of a scaled three-phase angle $1$ and an apex height of $\xiA=500$ on top of a precursor film of $\xiP=1$. 
The introduction of a precursor avoids the three-phase line singularity.
A series of simulations with different apex heights on the same precursor showed that for $\xiA\gtrapprox 100$, the coalescence behavior is no longer significantly influenced by the presence of the precursor.
The large height ratio of $500$ between drops and precursor warrants realistic conditions~\citep{Borcia:PhysRevE86}.
This also implies that the ``microscopic contact angle'' (determined by the slope in the inflection point near the contact line), is virtually identical to the overall aspect ratio (determined by the apex height over the footprint width) of the profile.
Details on the simulation method can be found in the supplemental material.


\begin{figure}
	\begin{center}%
		\includegraphics[width=0.96\colwidth]{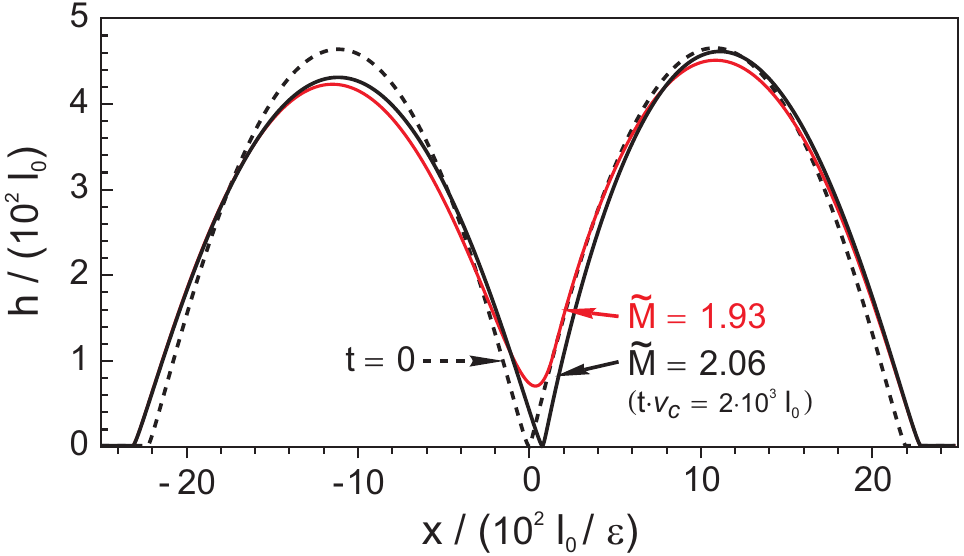}%
	\end{center}%
	\caption{\label{fig:cw:theo1}Snapshots of drop profiles during numerical simulations (see supplement for details). Dashed: moment of drop-drop contact ($t=0$). Solid: later on, at $t\cdot v_c= 2\cdot 10^{3} \lzero$, for $\M<\Mcrit$ (red) and $\M>\Mcrit$ (black). With $\lzero = \unit[5\cdot 10^{-7}]{m}$, the apex heights match those from the experiments.}
\end{figure}

Fig.~\ref{fig:cw:theo1} shows simulated drop profiles with $\Deff=0.125$ and two different $\M$.
Profiles are shown at the moment of contact ($\tau=0$) and after contact at $\tau=20$.
The profiles at contact are nearly identical for both $\M$ because $\M$ has only a weak impact on the spreading behavior on the thin precursor.
After contact both cases develop very differently.
For $\M=1.95$, the neck height $h_N$ grows, whereas for $\M=2.12$ it remains constantly low.
This means immediate coalescence and temporary noncoalescence, respectively, with $\Mcrit\approx2$.

\begin{figure}
	\begin{center}%
		\includegraphics[width=\colwidth]{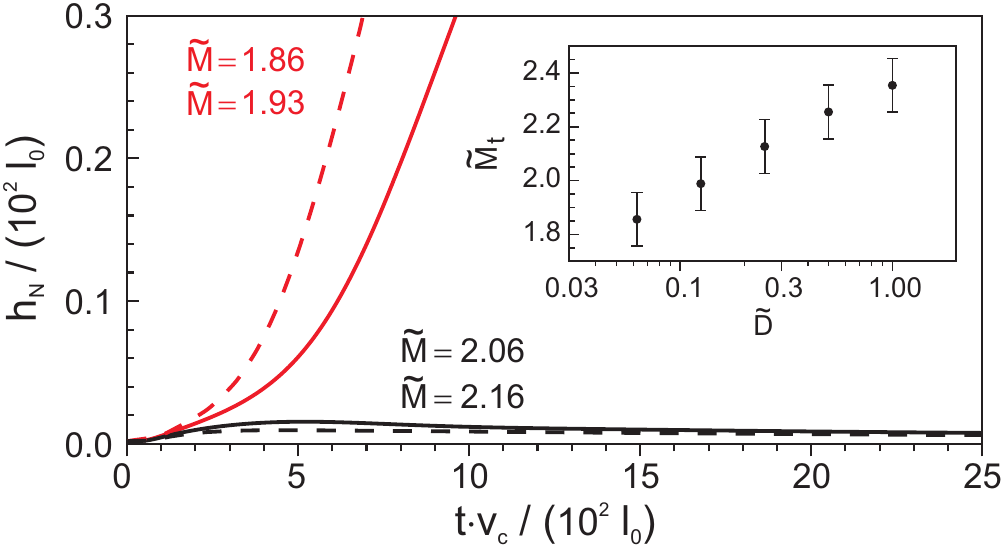}%
	\end{center}%
	\caption{\label{fig:cw:theo2}Neck height $\hN$ vs. time for $\M<\Mcrit$ (red) and $\M>\Mcrit$ (black). The solid curves correspond to the profiles from Fig.~\ref{fig:cw:theo1}. The inset shows the threshold Marangoni number $\Mcrit$ as a function of the effective diffusion constant $\Deff$.}
\end{figure}

The simulations indicate that $\Mcrit$ weakly depends on $\Deff$ ($\propto\Deff^{\nicefrac{1}{10}}$ for $5\cdot 10^{-2}\leq\Deff\leq 1$, see inset of Fig.~\ref{fig:cw:theo2}).
The simulation data nicely reproduce the experimental findings depicted in Figures~\ref{fig:cw:collage} and \ref{fig:cw:raw}.
In particular, for $\Deff=0.125$ they simulate systems with the same $\Mcrit\approx2$.
The simulations also propose a rather sharp transition (within $\pm \unit[5]{\%}$ of $\Mcrit$) between the two coalescence regimes for this range of $\Deff$.

How do these findings translate into physical dimensions?
For the experiments, $D$ and its scaled equivalent $\Deff$, are not known exactly.
Considering the drops in Fig.~\ref{fig:cw:collage}, $\lzero=\unit[5\cdot 10^{-7}]m$ (matching $\xiA=500$ to $\hA=\unit[2.5\cdot10^{-4}]{m}$), $\ThetaAdjAv=10\degrees$ (i.e., $\epsilon=0.175$), $\eta\ =\unit[2.5\cdot10^{-3}]Pa\cdot s$, and $\gAv\ =\unit[27\cdot10^{-3}]N/m$.
With these values, $v_c = \unit[3.5\cdot 10^{-3}]{m/s}$, and $\Deff=0.125$ translates into $D=\unit[6.5\cdot10^{-9}]m^2/s$.
This is a reasonable value and corroborates the validity of the simulations~\citep[see][]{Cussler:Diffusion}.
Translating the transition range of $\approx\pm\unit[5]{\%}$ of $\Mcrit$ into contact angles for $\Dg = \unit[0.87]{N/m}$ yields a transition within $\approx\pm0.2\degrees$, which is below a reasonable measurement resolution.

A disjoining pressure would act in the range of $\lessapprox\unit[10]{nm}$, which translates into $2\cdot 10^{-2} l_0$. The neck height at the decision between coalescence and noncoalescence ($t\cdot v_c \approx 2\cdot 10^2 l_0$ in Fig.~\ref{fig:cw:theo2}) is already around $2 l_0$. Thus we estimate that the disjoining pressure is not relevant for the decision, which is supported by the quantitative agreement of our simulations (without disjoining pressure) and the experimental threshold $\Mcrit$. Rather, it seems important to use a large height ratio between the drop apices and the precursor in the simulations.


\section{Summary and Conclusions}%
We present the first comprehensive experimental results on a sharp transition in coalescence behavior of sessile drops from different but completely miscible liquids. 
A wide range of three-phase angles was investigated for different $\gamma_1$, $\gamma_2$, $\Dg$ and liquid viscosities $\eta$.
The average three-phase angle $\ThetaAdjAv$ has been identified as main topographic control parameter.
The transition between the regimes of immediate coalescence and noncoalescence is sharp with respect to the threshold three-phase angle $\ThetaAdjAvCrit\propto\sqrt{\Dg/\gAv}$. 
For $\ThetaAdjAv >\ThetaAdjAvCrit$ coalescence occurs immediately, for $\ThetaAdjAv<\ThetaAdjAvCrit$ (temporary) noncoalescence is observed.

From the scaling of the lubrication equation we extract a characteristic Marangoni number $\M=3\Dg/(2\gAv(\ThetaAdjAv\cdot\pi/180\deg)^2)$ as measure for the competition between the coalescence-promoting capillarity and the drop-separating Marangoni flow. 
It is the key system parameter regarding the coalescence behavior and unifies the impact of the physicochemical liquid properties and the shape of the liquid bodies in the contact region. 
The experimental data for the threshold $\ThetaAdjAvCrit$ collapse into a single threshold Marangoni number $\Mcrit\approx 2$. 
This is in quantitative agreement with simulations assuming a realistic value for the diffusivity. 
The simulations also reveal a rather sharp border and that $\Mcrit$ depends on the diffusivity of liquid 2 in liquid 1.

The concepts and results are relevant in numerous cases, whenever two bulk liquid bodies (drops, films) of different composition get in contact.


\acknowledgments{We acknowledge helpful discussions with Jacco Snoeijer, Uwe Thiele, and Michael K\"opf, and technical support from Ferec Liebig. We also thank H. M\"ohwald for scientific advice and general support. S.K. was supported by the DFG (RI529/16-1) and LAM Research AG, Austria}.
%
%
%
\bibliographystyle{jfm}

%
%
%
%
\end{document}